\documentclass[12pt]{iopart}
\begin{document}

\title[On some class of reductions for Itoh-Narita-Bogoyavlenskii lattice]{On some class of reductions for Itoh-Narita-Bogoyavlenskii lattice}

\author{A K Svinin}

\address{Institute for System
Dynamics and Control Theory, Siberian Branch of
Russian Academy of Sciences, Russia}
\ead{svinin@icc.ru}
\begin{abstract}
We show a broad class of constraints compatible with Itoh-Narita-Bogoyavlenskii lattice hierarchy. All these constraints can be written in the form of discrete conservation law $I_{i+1}=I_i$ with appropriate homogeneous polynomial discrete function $I=I[a]$.
\end{abstract}

\pacs{02.30.Ik}

\noindent{\it Keywords}: KP hierarchy, Itoh-Narita-Bogoyavlenskii lattice


\section{Introduction}
\label{sec:I}
The aim of this paper is to show explicitly some class of constraints compatible with extended Volterra lattice
\begin{equation}
a_i^{\prime}=a_i\left(\sum_{j=1}^na_{i-j}-\sum_{j=1}^na_{i+j}\right),
\label{Bl}
\end{equation}
which we consider as a single evolution equation on the unknown function $a_i\equiv a(i, x)$ of discrete variable $i\in{\bf Z}$ and continuous variable $x\in{\bf R}$. For any $n\geq 1$, this equation is known to be integrable discretization for Kortweg-de Vries equation \cite{bogoyavlenskii}. Narita in the work \cite{narita}, making of use Hirota's method, showed that extended Volterra lattice admits soliton solutions. In \cite{itoh1} Itoh considered Lotka-Volterra systems which are equivalent to equation (\ref{Bl}) supplemented by specific periodicity condition $a_{i+2n+1}=a_i$.
In what follows we call equation (\ref{Bl}) Itoh-Narita-Bogoyavlenskii (INB) lattice. Perhaps the most interesting case from the point of applications is $n=1$, corresponding to Volterra lattice \cite{volterra}, \cite{kac}.

Equation (\ref{Bl}) is known to admit the hierarchy of pair-wise commuting generalized symmetries which, as is shown in the paper, can be written in the form
\begin{equation}
\partial_sa_i=(-1)^sa_i\left(S_{sn-1}^{(n, s)}(i-(s-1)n+1)-S_{sn-1}^{(n, s)}(i-sn)\right),
\label{Bli}
\end{equation}
where $\partial_s$ stands for derivative with respect to evolution parameter $t_s$ with $s\geq 2$. The functions $S_s^{(n, l)}[a]$ will be explicitly defined in Section \ref{sec:nb}. The INB equation itself can be written in the form (\ref{Bli}) with $t_1=x$ and $S_{n-1}^{(n, 1)}=\sum_{j=1}^na_{i+j-1}$. When the hierarchy is represented as (\ref{Bli}), it is clear that stationarity condition attached to some evolution parameter $t_s$, can be written as periodicity condition $S_{sn-1}^{(n, s)}(i+n+1)=S_{sn-1}^{(n, s)}(i)$.
It turns out that there exist a wide class of homogeneous polynomial discrete functions $I=I[a]$ defined by\footnote{Here $J\subset{\bf Z}^{l+1}$ is some finite indexing set.}
\[
I_i=\sum_{(j_1,\ldots, j_{l+1})\in J}a_{i+j_1}\cdots a_{i+j_{l+1}}
\]
for which periodicity condition $I_{i+T}=I_i$, with respective period $T\in{\bf Z}$ is consistent with INB lattice and its hierarchy. The set of all invariant constraints considered in the paper naturally include periodicity conditions $a_{i+T}=a_i$.

We consider equation (\ref{Bl}) and its hierarchy as a simplest case of reduction of the so-called Darboux-KP (DKP) chain hierarchy which in fact is a bi-infinite sequence of KP hierarchies and our main goal, in fact, is to  approve our previous results \cite{svinin4} on this simple example. For completeness, we give, in the next section, preliminaries on our approach to investigate some class of integrable lattices related to the KP hierarchy. In Section \ref{sec:nb}, we show compatible constraints for equation (\ref{Bl}) in its explicit form in Theorem 3. The Section 4 is devoted to Volterra lattice and its reductions. We write down, in this section, attached systems of ordinary differential equations generated by corresponding constraints and its discrete symmetry transformations. Also we claim the relations defining spectral curves associated with Lax matrices.

\section{Preliminaries on DKP chain hierarchy and its invariant submanifolds}

\subsection{DKP chain hierarchy}
In \cite{svinin1} and \cite{svinin2} we have developed an approach in which a broad community of integrable differential-difference equations (lattices) are related to KP hierarchy. In a paper \cite{svinin4} we have shown that these integrable lattices admit a wide class of constraints compatible with all higher flows of its hierarchy. An objective of this section is to provide the reader by information about the DKP chain hierarchy and its reductions.

Integrable lattices in our geometric set-up naturally appear as a result of reductions of a bi-infinite sequence of KP hierarchies whose equations of motion we write in the form
of two evolution generating equations \cite{mpz}
\begin{eqnarray}
\partial_sh(i)&=&\partial H^{(s)}(i),
\label{1}
\\
\partial_sa(i)&=&a(i)\left(H^{(s)}(i+1)-H^{(s)}(i)\right).
\label{2}
\end{eqnarray}
The first relation (\ref{1}) yields evolution equations of KP hierarchy in the form of local conservation laws \cite{wilson}. Laurent series: generating functions for conserved densities and corresponding fluxes of KP hierarchy
\[
h(i)=
z+\sum_{k\geq 2}h_k(i)z^{-k+1}\;\;\;\mbox{and}\;\;\;
H^{(s)}(i)=
z^s+\sum_{k\geq 1}H_k^s(i)z^{-k},\;\;
\]
are related with KP wave functions 
\[
\psi_i=\left(1+\sum_{k\geq 1}w_k(i)z^{-k}\right)\exp(\sum_{s\geq 1}t_sz^s)
\] 
as 
\[
h(i)=\partial\psi_i\cdot\psi_i^{-1}\;\;\;\mbox{and}\;\;\; 
H^{(s)}(i)=\partial_s\psi_i\cdot\psi_i^{-1},
\]
respectively. In turn, the Laurent series $a(i)=z+\sum_{k\geq 1}a_k(i)z^{-k+1}$ is calculated as $a(i)=z\psi_{i+1}\cdot\psi_i^{-1}$. We call equations (\ref{1}) and (\ref{2}) DKP chain hierarchy. It is useful to rewrite generating equation (\ref{2}) in the form of differential-difference conservation law
\[
\partial_s\xi(i)=H^{(s)}(i+1)-H^{(s)}(i)
\]
with
\begin{eqnarray}
\fl
\displaystyle
\xi(i)&=&\ln a(i)=\ln z +
\sum_{k\geq 1}a_k(i)z^{-k}-
\frac{1}{2}\left(\sum_{k\geq 1}a_k(i)z^{-k}\right)^2
+\frac{1}{3}\left(\sum_{k\geq 1}a_k(i)z^{-k}\right)^3-\cdots
\nonumber \\
\fl
&\equiv&\ln z+\sum_{k\geq 1}\xi_k(i)z^{-k}. \nonumber
\end{eqnarray}
Thus, more exactly, equations (\ref{1}) and (\ref{2}) can be written as follows:
\[
\partial_sh_k(i)=\partial H^s_{k-1}(i),\;\;\;
\partial_s\xi_k(i)=H^s_k(i+1)-H^s_k(i).
\]

To establish relationship of integrable lattices like INB lattice (\ref{Bl}), Shabat dressing lattice \cite{shabat}, Toda lattice \cite{toda}, Belov-Chaltikian lattice \cite{belov} and so on, with KP hierarchy, the following two theorems are useful.

{\bf Theorem 1.} {\it
\cite{svinin1} The submanifold ${\cal S}_{l-1}^n$ defined by condition
\begin{equation}
z^{l-n}a^{[n]}(i)\in{\cal H}_{+}(i),\;\; \forall
i\in{\bf Z} \label{sln}
\end{equation}
is tangent with respect to DKP chain flows defined by (\ref{1}) and (\ref{2}).      }

{\bf Theorem 2.} {\it
\cite{svinin2} The chain of inclusions
of invariant submanifolds
\[
{\cal S}_{l-1}^n\subset {\cal S}_{2l-1}^{2n}\subset{\cal
S}_{3l-1}^{3n}\subset\cdot\cdot\cdot\subset {\cal
S}_{kl-1}^{kn}\subset\cdot\cdot\cdot
\]
is valid.          }

\noindent
Here, by definition
\[
a^{[s]}(i)=\left\{
\begin{array}{l}
\prod_{j=1}^sa(i+j-1),\;\;\; s\geq 1 \\ 1,\;\;\; s=0  \\
\prod_{j=1}^{|s|}a^{-1}(i-j),\;\;\; s\leq -1
\end{array}
\right.
\]
are discrete Fa\`a di Bruno iterates of Laurent series $a(i)$. It is obvious that the coefficients $a_j^{[s]}$ defined through the relation\footnote{We use simplified notations $a^{[s]}\equiv a^{[s]}(i)$, $a_j^{[s]}\equiv a_j^{[s]}(i)$ in formulas which contain no shifts with respect to discrete variable $i\in{\bf Z}$.}
\[
a^{[s]}=z^s+\sum_{j\geq 1}a_j^{[s]}z^{s-j}
\] 
are the discrete functions of $(a_1,\ldots, a_j)$. These functions are related with each other by obvious relation
\begin{equation}
\fl
a_k^{[s_1+s_2]}(i)=a_k^{[s_1]}(i)+\sum_{j=1}^{k-1}a_j^{[s_1]}(i)a_{k-j}^{[s_2]}(i+s_1)+a_k^{[s_2]}(i+s_1)=(s_1\leftrightarrow s_2).
\label{s1s2}
\end{equation}
Here the symbol $(s_1\leftrightarrow s_2)$ denotes the same right-hand side of this relation but with mutually replaced $s_1$ and $s_2$. This relation will be extensively used throughout the paper.

We observe that the condition (\ref{sln}) can be written in the form of the following generating relation:
\[
z^{l-n}a^{[n]}=H^{(l)}+\sum_{k=1}^la_k^{[n]}H^{(l-k)}.
\]

\subsection{$n$-th discrete KP hierarchy and its reductions}
When restricting DKP chain hierarchy on ${\cal S}_0^n$, all the coefficients $H_k^s$ become discrete polynomial functions of $(a_1,\ldots, a_{k+s})$ defined by \cite{svinin1}, \cite{svinin4}
\begin{equation}
H^s_k=F_k^{(n,s)}\equiv a_{k+s}^{[sn]}+\sum_{j=1}^{s-1}q_j^{(n, sn)}a_{k+s-j}^{[(s-j)n]},
\label{fkns}
\end{equation}
where $q_k^{(n,r)}=q_k^{(n,r)}[a_1, a_2,\ldots, a_k]$, by definition, are polynomial discrete functions defined through generating relation
\begin{equation}
z^r=a^{[r]}+\sum_{j\geq 1}q_j^{(n,r)}z^{j(n-1)}a^{[r-jn]}
\label{zr}
\end{equation}
which yields
\begin{equation}
a_k^{[r]}+\sum_{j=1}^{k-1}a^{[r-jn]}_{k-j}q_j^{(n,r)}+q_k^{(n,r)}=0,\;\;\;
\forall k\geq 1.
\label{exact}
\end{equation}
Let us write down the first few $q_k^{(n,r)}$
\[
q_1^{(n, r)}=-a_1^{[r]},\;\;\; q_2^{(n, r)}=-a_2^{[r]}+a_1^{[r]}a_1^{[r-n]},\;\;\;
\]
\[
q_3^{(n, r)}=-a_3^{[r]}+a_1^{[r]}a_2^{[r-n]}+a_1^{[r-2n]}a_2^{[r]}-a_1^{[r]}a_1^{[r-n]}a_1^{[r-2n]}.
\]
It can be checked that a more general relation than (\ref{exact}), namely \cite{svinin4}
\begin{equation}
a_k^{[r]}(i)+\sum_{j=1}^{k-1}a_{k-j}^{[r-jn]}(i)q_j^{(n,r-p)}(i+p)+q_k^{(n,r-p)}(i+p)=a_k^{[p]}(i)
\label{integers}
\end{equation}
with any $p\in\textbf{Z}$ is valid. Solving this in favor of $q_k^{(n,r-p)}(i+p)$ yields
\[
a_k^{[p]}(i)+\sum_{j=1}^{k-1}q_j^{(n,r-(k-j)n)}(i)a_{k-j}^{[p]}(i)+q_k^{(n,r)}(i)=q_k^{(n,r-p)}(i+p).
\]

One sees that, when restricting on ${\cal S}_0^n$, DKP chain hierarchy is reduced to evolution equations in the form of differential-difference conservation law
\begin{equation}
\partial_s\xi_k(i)=F_k^{(n,s)}(i+1)-F_k^{(n,s)}(i),
\label{dKP}
\end{equation}
where $F_k^{(n,s)}$ is given by (\ref{fkns}). We refer to these equations with some fixed $n\geq 1$ as $n$-th discrete KP hierarchy. It is worth to notice that these equations, in fact, appear as a result of restriction of DKP chain hierarchy on ${\cal S}_0^n$ and subsequent projection of dynamics on the space ${\cal M}$ whose points are defined by infinite number of functions of discrete variable $(a_1, a_2,\ldots)$. One can say that ${\cal M}$ has infinite functional dimension. Let us denote by ${\cal M}_k$ the space whose points are defined by finite number $(a_1,\ldots a_k)$ of functions of discrete variable $i$ and $\pi_k : {\cal M}\mapsto{\cal M}_k$ being a natural projection. It is obvious that reduction of  DKP chain hierarchy on the intersection ${\cal S}_0^n\cap{\cal S}_{l-1}^p$ is equivalent to restriction of the flows given by (\ref{dKP}) on some submanifold
${\cal M}_{n, p, l}\subset{\cal M}$. This submanifold is defined by infinite number of algebraic equations \cite{svinin4}
\begin{equation}
J_k^{(n, p, l)}[a_1,\ldots, a_{k+l}]=0,\;\;\; k\geq 1
\label{alg}
\end{equation}
with
\[
J_k^{(n, p, l)}(i)=a^{[p]}_{k+l}(i)-a_{k+l}^{[ln]}(i)-\sum_{j=1}^{l-1}q_j^{(n,ln-p)}(i+p)a_{k+l-j}^{[(l-j)n]}(i).
\]
Observe, that in the case $p=ln$ the relations $J_k^{(n, ln, l)}=0$ are identities and therefore produce no nontrivial submanifold of ${\cal M}$.
This is so because ${\cal S}_0^n\subset{\cal S}_{l-1}^{ln}$ thanks to the Theorem 2.

Taking into account (\ref{integers}) we can also write
\begin{equation}
J_k^{(n, p, l)}(i)=q_{k+l}^{(n,ln-p)}(i+p)+\sum_{j=1}^{k-1}a_j^{[-(k-j)n]}(i)q_{k+l-j}^{(n,ln-p)}(i+p).
\label{jk}
\end{equation}
Let us denote $Q_k^{(n, p, l)}(i)=q_{k+l}^{(n,ln-p)}(i+p)$ and therefore one can rewrite the relation (\ref{jk}) as
\begin{equation}
J_k^{(n, p, l)}=Q_k^{(n, p, l)}+\sum_{j=1}^{k-1}a_j^{[-(k-j)n]}Q_{k-j}^{(n, p, l)}.
\label{jk1}
\end{equation}
It is evident that the submanifold ${\cal M}_{n, p, l}$ can be equivalently defined by algebraic equations
$Q_k^{(n, p, l)}[a_1,\ldots, a_{k+l}]=0$. Solving (\ref{jk1}) in favor of $Q_k^{(n, p, l)}$ yields
\begin{equation}
Q_k^{(n, p, l)}=J_k^{(n, p, l)}+\sum_{j=1}^{k-1}q_j^{(n,-(k-j)n)}J_{k-j}^{(n, p, l)}.
\label{qj}
\end{equation}
By definition, one has $Q_k^{(n, p+n, l+1)}(i)=Q_{k+1}^{(n, p, l)}(i+n)$. Making of use this relation, (\ref{integers}) and (\ref{qj}), one can prove that
\begin{equation}
J_k^{(n, p+n, l+1)}(i)=J_{k+1}^{(n, p, l)}(i+n)+\sum_{j=1}^ka_j^{[n]}(i)J_{k-j+1}^{(n, p, l)}(i+n).
\label{rec}
\end{equation}
For $Q_k^{(n, p, l)}$ we are able to write down following evolution equations
\begin{eqnarray}
D_{t_s}Q_k^{(n, p, l)}(i)&=&
Q_{k+s}^{(n, p, l)}(i+sn)+\sum_{j=1}^sq_j^{(n, sn)}(i+p)Q_{k+s-j}^{(n, p, l)}(i+(s-j)n)
\nonumber \\
&&-Q_{k+s}^{(n, p, l)}(i)-\sum_{j=1}^sq_j^{(n, sn)}(i-(s+k-j)n)Q_{k+s-j}^{(n, p, l)}(i)
\nonumber
\end{eqnarray}
which easily derived from
\begin{eqnarray}
D_{t_s}q_k^{(n, r)}(i)&=&q_{k+s}^{(n, r)}(i+sn)+\sum_{j=1}^sq_j^{(n,sn)}(i)q_{k+s-j}^{(n, r)}(i+(s-j)n) \nonumber \\
& &-q_{k+s}^{(n, r)}(i)-\sum_{j=1}^sq_j^{(n,sn)}(i+r-(k+s-j)n)q_{k+s-j}^{(n, r)}(i). \label{dts}
\end{eqnarray}
The latter, in turn, comes from Lax equation attached to linear problem for wave function $\Psi=\{\psi_i : i\in{\bf Z}\}$ coded in (\ref{zr}) \cite{svinin}.

As was shown in \cite{svinin4}, there exist weaker than (\ref{alg}) conditions invariant with respect to $n$-th discrete KP hierarchy equations (\ref{dKP}). Desired constraints are written as periodicity conditions
\begin{equation}
I_k^{(n, p, l)}(i+n)=I_k^{(n, p, l)}(i)
\label{pc}
\end{equation}
with
\begin{eqnarray}
I_k^{(n, p, l)}(i)&=&J_k^{(n, p, l)}(i)+\sum_{j=1}^{k-1}a_j^{[-p]}(i+p)J_{k-j}^{(n, p, l)}(i) \nonumber \\
                  &=&Q_k^{(n, p, l)}(i)+\sum_{j=1}^{k-1}a_j^{[-p-(k-j)n)]}(i+p)Q_{k-j}^{(n, p, l)}(i). \nonumber
\end{eqnarray}
Moreover, if the conditions (\ref{pc}) are valid then $I_k^{(n, p, l)}$ do not depend on evolution parameters $t_s$, i.e., $D_{t_s}I_k^{(n, p, l)}\equiv 0,\;\;\; \forall s\geq 1$. For these functions, we can write
\[
I_k^{(n, p+n, l+1)}(i)=I_{k+1}^{(n, p, l)}(i+n)-a_k^{[-p-n]}(i+p+n)I_1^{(n, p, l)}(i+n)
\]
in parallel with (\ref{rec}). It can be shown that relationship of $I_k^{(n, p, l)}$'s with wave KP functions is given by
\begin{equation}
\sum_{j\geq 1}I_j^{(n, p, l)}(i)z^{-j}=z^l-\frac{1}{\psi_{i+p}}\left(z^l\psi_{i+ln}+\sum_{j=1}^lz^{l-j}q_j^{(n, ln-p)}(i+p)\psi_{i+(l-j)n}\right)
\label{ikz}
\end{equation}

If $d$ any divisor of $n$, then the set of conditions $I_k^{(n, p, l)}(i+d)=I_k^{(n, p, l)}(i)$ also produce invariant submanifold of $n$-th discrete KP hierarchy. We denote corresponding submanifold by ${\cal N}_{n, p, l}^d$. It is evident that ${\cal M}_{n, p, l}\subset{\cal N}_{n, p, l}^d\subset{\cal N}_{n, p, l}^n$.

In the following section we consider invariant constraints for Narita-Bogoyavlenskii lattice which correspond to restriction of $n$-th discrete KP hierarchy on ${\cal M}_{n, n+1, 1}$. Our aim to write invariant conditions appeared as a result of intersection of ${\cal M}_{n, n+1, 1}$ and ${\cal N}_{n, p, l}^d$ in its explicit form.

\section{Reductions for INB lattice}
\label{sec:nb}

\subsection{Restriction of $n$-th discrete KP hierarchy on ${\cal M}_{n, n+1, 1}$. INB lattice}

Let us consider invariant submanifold ${\cal M}_{n, n+1, 1}$ of phase-space ${\cal M}$ defined by algebraic equations
\begin{equation}
J_k^{(n, n+1, 1)}=-J_k^{(n+1, n, 1)}=a^{[n+1]}_{k+1}-a_{k+1}^{[n]}=0,\;\;\; k\geq 1
\label{3}
\end{equation}
for some fixed positive integer $n$. Taking into account (\ref{s1s2}) we can rewrite (\ref{3}) as
\begin{equation}
\sum_{j=1}^{k-1}a_{k-j}^{[n]}(i)a_j(i+n)+a_k(i+n)=0.
\label{4}
\end{equation}
One can easily check that these equations are solved by
\begin{equation}
a_k(i)=a_{k-1}^{[-n]}(i)a_i
\label{40}
\end{equation}
where $a_i\equiv a_1(i)$. Indeed, substituting the latter in (\ref{4}) we have
\[
a_{i+n}\left(a_{k-1}^{[n]}(i)+\sum_{j=1}^{k-2}a_{k-j-1}^{[n]}(i)a_j^{[-n]}(i+n)+a_{k-1}^{[-n]}(i+n)\right)
\]
\[
=a_{i+n}a_{k-1}^{[0]}(i)= 0.
\]
Here we have used (\ref{s1s2}). Following technical proposition is valid.

{\bf Proposition 1.} {\it
In virtue of relations (\ref{3})
\begin{equation}
a_k^{[s]}(i)=\sum_{j=1}^sa_{k-1}^{[-n+j-1]}(i)a_{i+j-1},\;\;\; \mbox{for $s\geq 1$}
\label{5}
\end{equation}
and
\[
a_k^{[s]}(i)=-\sum_{j=1}^{|s|}a_{k-1}^{[-n-j]}(i)a_{i-j},\;\;\; \mbox{for $s\leq -1$}.
\]
}

{\bf Proof.} Taking into account (\ref{s1s2}) and (\ref{40}), one has
\begin{eqnarray}
\fl
a_k^{[s+1]}(i)&=&a_k^{[s]}(i)+\sum_{j=1}^{k-1}a_{k-j}^{[s]}(i)a_j(i+s)+a_k(i+s) \nonumber \\
\fl
\displaystyle
&=&a_k^{[s]}(i)+a_{i+s}\left(a_{k-1}^{[s]}(i)+\sum_{j=1}^{k-2}a_{k-j-1}^{[s]}(i)a_j^{[-n]}(i+s)+a_{k-1}^{[-n]}(i+s)\right) \nonumber \\
\fl
&=&a_k^{[s]}(i)+a_{k-1}^{[-n+s]}(i)a_{i+s}. \nonumber
\end{eqnarray}
Making of use this formula one can successively to prove (\ref{5}) for $k=2, 3,...$ by induction with respect to $s$.

With (\ref{5}), we are able to calculate all $a_k$ as discrete functions of $a$ to obtain
\[
a_2(i)=-a_i\sum_{j=1}^na_{i-j},\;\;\;
a_3(i)=a_i\sum_{j_1=1}^na_{i-j_1}\left(\sum_{j_2=1}^{n+j_1}a_{i-j_2}\right),\;\;\;
\]
\[
a_4(i)=-a_i\sum_{j_1=1}^na_{i-j_1}\left(\sum_{j_2=1}^{n+j_1}a_{i-j_2}\left(\sum_{j_3=1}^{n+j_2}a_{i-j_3}\right)\right)
\]
and so on.

What we learn from the above calculations is that the restriction of $n$-th discrete KP hierarchy on ${\cal M}_{n, n+1, 1}$ and subsequent projection $\pi_1 : {\cal M}\mapsto{\cal M}_1$ generate the hierarchy of evolution equations in the form of differential-difference conservation laws
\[
\partial_sa_i=F_1^{(n, s)}(i+1)-F_1^{(n, s)}(i)
\]
together with conservation laws  (\ref{dKP}), where conserved densities $\xi_k=\xi_k[a]$ and fluxes $F_k^{(n, s)}[a]$ are some homogeneous polynomials of $k$-th and $(k+s)$-th power, respectively. For the first flow we have
\begin{eqnarray}
a_i^{\prime}&=&F_1^{(n, 1)}(i+1)-F_1^{(n, 1)}(i) \nonumber \\
            &=&a_2^{[n]}(i+1)-a_2^{[n]}(i).       \nonumber
\end{eqnarray}
To calculate the right-hand side of this equation as discrete function of $a=a_i$, it is convenient to use
\begin{eqnarray}
a_2^{[n+1]}(i)&=&a_2^{[n]}(i)+a_1^{[n]}(i)a_{i+n}+a_2(i+n) \nonumber \\
&=&a_2^{[n]}(i+1)+a_1^{[n]}(i+1)a_i+a_2(i), \nonumber
\end{eqnarray}
where $a_2(i)=a_1^{[-n]}(i)a_i=-a_1^{[n]}(i-n)a_i$. Taking this into account, we get
\begin{eqnarray}
a_i^{\prime}&=&a_i\left(a_1^{[n]}(i-n)-a_1^{[n]}(i+1)\right) \nonumber \\
            &=&a_i\left(\sum_{j=1}^na_{i-j}-\sum_{j=1}^na_{i+j}\right),       \nonumber
\end{eqnarray}
which is nothing but INB equation (\ref{Bl}).

\subsection{The functions $S^{(n, l)}_s[a]$ and $T^{(n, l)}_s[a]$ and its properties}

Let us prepare, for further use, the discrete functions $S^{(n, l)}_s[a]$ and $T^{(n, l)}_s[a]$  through
\begin{equation}
S^{(n, l)}_s(i)=\sum_{0\leq j_{l-1}\leq\cdots\leq j_0\leq s}\left(\prod_{k=0}^{l-1}a_{i+kn+j_k}\right)
\label{snls}
\end{equation}
with $l\geq 0$ and $s\geq 0$ and
\[
T^{(n, l)}_s(i)=\sum_{0\leq j_0<\cdots <j_{l-1}\leq s}\left(\prod_{k=0}^{l-1}a_{i+kn+j_k}\right)
\]
with $l\geq 0$ and $s\geq l-1$. We observe that the functions $S^{(n, l)}_s$ satisfy following relations:
\begin{equation}
S^{(n, l)}_s(i)-S^{(n, l)}_{s-d}(i)=\sum_{j=1}^da_{i+s-j+1}S^{(n, l-1)}_{s-j+1}(i+n),
\label{s1}
\end{equation}
\begin{equation}
S^{(n, l)}_s(i)-S^{(n, l)}_{s-d}(i+d)=\sum_{j=1}^da_{i+(l-1)n+j-1}S^{(n, l-1)}_{s-j+1}(i+j-1)
\label{s2}
\end{equation}
for $d=1,\ldots, s$ and
\begin{eqnarray}
S^{(n, l)}_s(i)&=&\sum_{j=1}^{s+1}a_{i+s-j+1}S^{(n, l-1)}_{s-j+1}(i+n) \nonumber \\
               &=&\sum_{j=1}^{s+1}a_{i+(l-1)n+j-1}S^{(n, l-1)}_{s-j+1}(i+j-1). \nonumber
\end{eqnarray}
For $T^{(n, l)}_s$ we have the identities
\begin{equation}
T^{(n, l)}_s(i)-T^{(n, l)}_{s-d}(i+d)=\sum_{j=1}^da_{i+j-1}T^{(n, l-1)}_{s-j}(i+n+j),
\label{t1}
\end{equation}
\begin{equation}
T^{(n, l)}_s(i)-T^{(n, l)}_{s-d}(i)=\sum_{j=1}^da_{i+(l-1)n+s-j+1}T^{(n, l-1)}_{s-j}(i)
\label{t2}
\end{equation}
with $d=1,\ldots, s-l+1$ and
\begin{eqnarray}
T^{(n, l)}_s(i)&=&\sum_{j=1}^{s-l+2}a_{i+j-1}T^{(n, l-1)}_{s-j}(i+n+j) \label{tnl1} \\
        &=&\sum_{j=1}^{s-l+2}a_{i+(l-1)n+s-j+1}T^{(n, l-1)}_{s-j}(i). \label{tnl}
\end{eqnarray}

\subsection{A class of reductions for INB lattice}

With $S^{(n, l)}_s$ and $T^{(n, l)}_s$ in hand, we are in position to prove

{\bf Proposition 2.} {\it On ${\cal M}_{n, n+1, 1}$
\begin{equation}
\fl
J_k^{(n, ln+s+1, l)}(i)=T^{(n, k+l)}_s(i-(k-1)n)+\sum_{j=1}^{k-1}a_j^{[-(k-j)n]}(i)T^{(n, k+l-j)}_s(i-(k-j-1)n)
\label{Jk1}
\end{equation}
for $s\geq l$ and $J_k^{(n, ln+s+1, l)}\equiv 0$ for $s=0,\ldots, l-1$ and
\begin{equation}
\fl
J_k^{(n, ln-s-1, l)}(i)=R^{(n, k+l)}_s(i-(k-1)n)+\sum_{j=1}^{k-1}a_j^{[-(k-j)n]}(i)R^{(n, k+l-j)}_s(i-(k-j-1)n)
\label{Jk2}
\end{equation}
for $s\geq 0$, where $R^{(n, k)}_s(i)\equiv (-1)^kS^{(n, k)}_s(i-s-1)$.
}

Let us give some remarks. It is accepted in (\ref{Jk1}), that $T^{(n, k)}_s\equiv 0$ with $s\leq k-2$. For example
$J_k^{(n, ln+l+1, l)}=a_{k-1}^{[-n]}T^{(n, l+1)}_l$. As a corollary of this proposition, one has ${\cal M}_{n, n+1, 1}\subset{\cal M}_{n, ln+s+1, l}$,
with $l\geq 1$ and $s=0,\ldots, l-1$. We observe comparing (\ref{Jk1}) and (\ref{Jk2}) with (\ref{jk}) that this proposition can be reformulated in the following equivalent form.

{\bf Proposition 3.} {\it On ${\cal M}_{n, n+1, 1}$
\[
q_k^{(n, s+1)}(i)=(-1)^kS_s^{(n, k)}(i-(k-1)n)
\]
for $s\geq 0$ and
\[
q_k^{(n, -s-1)}(i)=T_s^{(n, k)}(i-(k-1)n-s-1)
\]
for $s\geq k-1$ and $q_k^{(n, -s-1)}\equiv 0$ for $s=0,\ldots, k-2$.
}

From this proposition and the relation \cite{svinin}
\[
\fl
q_k^{(n, s_1+s_2)}(i)=
q_k^{(n, s_1)}(i)+\sum_{j=1}^{k-1}q_j^{(n, s_1)}(i)q_{k-j}^{(n, s_2)}(i+s_1-jn)+q_k^{(n, s_2)}(i+s_1)=(s_1\leftrightarrow s_2)
\]
we get two identities
\[
S^{(n, l)}_s(i)+\sum_{j=1}^{l-1}(-1)^jS^{(n, l-j)}_s(i)T^{(n, j)}_s(i+(l-j)n)+(-1)^lT^{(n, l)}_s(i)=0
\]
and
\[
T^{(n, l)}_s(i)+\sum_{j=1}^{l-1}(-1)^jT^{(n, l-j)}_s(i)S^{(n, j)}_s(i+(l-j)n)+(-1)^lS^{(n, l)}_s(i)=0
\]
establishing the relationship between the discrete functions $S^{(n, l)}_s[a]$ and $T^{(n, l)}_s[a]$.

{\bf Proof of Proposition 2.} To save the space we restrict ourselves by the sketch of the proof. We prove, by induction with respect to $k$, the validity of (\ref{Jk1}) for $l=0$, i.e. that on ${\cal M}_{n, n+1, 1}$
\begin{equation}
\fl
a_k^{[s+1]}(i)=T_s^{(n, k)}(i-(k-1)n)+\sum_{j=1}^{k-1}a_j^{[-(k-j)n]}(i)T^{(n, k-j)}_s(i-(k-j-1)n).
\label{aks1}
\end{equation}
In the case $k=1$ one has $a_1^{[s+1]}(i)=\sum_{j=0}^sa_{i+j}=T_s^{(n, 1)}(i)$, by definition. Suppose now that the relation (\ref{aks1}) is already proved for $k=1,\ldots ,k_0$. Then, for these values of $k$ and arbitrary $m\in\textbf{Z}$ we can show that
\[
\fl
a_k^{[m+s+1]}(i)-a_k^{[m]}(i)=
\]
\[
\fl
=T^{(n, k)}_s(i+m-(k-1)n)+\sum_{j=1}^{k-1}a_j^{[m-(k-j)n]}(i)T^{(n, k-j)}_s(i+m-(k-j-1)n).
\]
In particular
\begin{equation}
\fl
a_k^{[-n+s+1]}(i)-a_k^{[-n]}(i)=T^{(n, k)}_s(i-kn)+\sum_{j=1}^{k-1}a_j^{[-(k-j+1)n]}(i)T^{(n, k-j)}_s(i-(k-j)n).
\label{66}
\end{equation}
Taking into account (\ref{5}), we have
\[
a_{k+1}^{[s+1]}(i)-a_k^{[-n]}(i)T_s^{(n, 1)}(i)=\sum_{j=1}^s\left(a_k^{[-n+j]}(i)-a_k^{[-n]}(i)\right)a_{i+j}.
\]
Then taking into account (\ref{tnl}) and (\ref{66}) we get
\[
\fl
a_{k+1}^{[s+1]}(i)-a_k^{[-n]}(i)T_s^{(n, 1)}(i)=\sum_{j=1}^{k-1}a_{k-j}^{[-(j+1)n]}(i)T^{(n, j+1)}_s(i-jn)+T^{(n, k+1)}_s(i-kn).
\]
Therefore we prove that if (\ref{aks1}) is valid for $k=1,\ldots, k_0$, then it is true for $k=k_0+1$. Thus, by induction, the relation (\ref{Jk1}) is proven for $l=0$. For remaining values $l\geq 1$ the functions $J_k^{(n, ln+s+1, l)}$ are calculated with the help of recurrence relation (\ref{rec}). Similar reasonings are used to prove (\ref{Jk2}).

With Proposition 3 we can easily write equations of INB lattice hierarchy. To this aim, we use the fact that on ${\cal M}_{n, n+1, 1}$ one has
\[
Q_k^{(n, n+1, 1)}(i)=q_{k+1}^{(n, -1)}(i+n+1)=0,\;\;\; \forall k\geq 1.
\]
Then from (\ref{dts}) we have
\begin{equation}
\partial_sq_1^{(n, -1)}(i)=q_1^{(n, -1)}(i)\left(q_s^{(n, sn)}(i)-q_s^{(n, sn)}(i-n-1)\right),
\label{in}
\end{equation}
where $q_1^{(n, -1)}(i)=a_{i-1}$, by definition. According to Proposition 3, on ${\cal M}_{n, n+1, 1}$ one has $q_s^{(n,sn)}(i)=(-1)^sS_{sn-1}^s(i-(s-1)n)$. Substituting the latter in (\ref{in}) we obtain evolution equations of INB hierarchy (\ref{Bli}).


Now we would like to write down invariant constraints corresponding to submanifold ${\cal M}_{n, n+1, 1}\cap{\cal N}_{n,p,l}^d$, where $d$ is any divisor of $n$ or $n+1$. From Proposition 2, we know that $I_1^{(n, ln+s+1, l)}(i)=T^{(n, l+1)}_s(i),$ for $s\ge l$ and $I_1^{(n, ln-s-1, l)}(i)=R^{(n, l+1)}_s(i)=(-1)^{l+1}S^{(n, l+1)}_s(i-s-1)$ for $s\geq 0$. It is natural to require so as intersection ${\cal M}_{n, n+1, 1}\cap{\cal N}_{n,p,l}^d$ to be nontrivial. This requirement means that condition $I_1^{(n, p, l)}(i+d)=I_1^{(n, p, l)}(i)$ must guarantee that $I_k^{(n, p, l)}(i+d)=I_k^{(n, p, l)}(i)$ is identity for all $k\geq 2$. Unfortunately, we are able analyze this only for the case $d=1$.


{\bf Theorem 3.} {\it Each one of the constraints
\begin{equation}
T^{(n, l+1)}_s(i+1)=T^{(n, l+1)}_s(i),\;\;\; s\geq l
\label{t3}
\end{equation}
and
\begin{equation}
S^{(n, l+1)}_s(i+1)=S^{(n, l+1)}_s(i),\;\;\; s\geq 0.
\label{s3}
\end{equation}
is consistent with INB lattice hierarchy.}



From (\ref{t1}) and (\ref{t2}), with $d=1$, we see that the condition (\ref{t3}) can be rewritten as the relation
\begin{equation}
a_{i+ln+s+1}T^{(n, l)}_{s-1}(i+1)=a_iT^{(n, l)}_{s-1}(i+n+1).
\label{TNh1}
\end{equation}
For (\ref{s3}), taking into account (\ref{s1}) and (\ref{s2}), with $d=1$, we have the relation
\begin{equation}
a_{i+s+1}S^{(n, l)}_s(i+n+1)=a_{i+ln}S^{(n, l)}_s(i).
\label{SNh1}
\end{equation}

{\bf Proof of Theorem 3.} To prove theorem, there is a need only to show that intersection ${\cal M}_{n, n+1, 1}\cap{\cal N}_{n,p,l}^1$ is nontrivial. Let us consider the condition (\ref{t3}). We observe that on ${\cal M}_{n,n+1,1}$ homogeneous discrete polynomials $I^{(n,p ,l)}_k[a]$ are calculated with the help of recurrence relation
\begin{equation}
\fl
I^{(n,p,l)}_k(i)=T^{(n,l+k)}_s(i-(k-1)n)-\sum_{j=1}^{k-1}T_{p+(k-j)n-1}^{(n,j)}(i-(k-1)n)I^{(n,p, l)}_{k-j}(i).
\label{recur}
\end{equation}
Suppose we already proved that in virtue of (\ref{t3}) equation
\begin{equation}
I^{(n,p,l)}_j(i+1)=I^{(n,p,l)}_j(i)
\label{step}
\end{equation}
is valid for $j=1,\ldots, k-1$. Then the relation $I^{(n,p,l)}_k(i+1)=I^{(n,p,l)}_k(i)$ can be written as
\[
\fl
T^{(n,l+k)}_s(i-(k-1)n+1)-T^{(n,l+k)}_s(i-(k-1)n)
\]
\begin{equation}
\fl
-\sum_{j=1}^{k-1}\left(T_{p+(k-j)n-1}^{(n,j)}(i-(k-1)n+1)-T_{p+(k-j)n-1}^{(n,j)}(i-(k-1)n)\right)I^{(n,p,l)}_{k-j}(i)=0.
\label{Ik}
\end{equation}
Making of use the identities (\ref{t1}) and (\ref{t2}), we can rewrite equation (\ref{Ik}) as
\[
\fl
a_{i+p}\left(T^{(n,l+k-1)}_{s-1}(i-(k-1)n+1)
-I_{k-1}^{(n,p,l)}(i)   \right.
\]
\[
\left.    -\sum_{j=1}^{k-2}T_{p+(k-j-1)n-2}^{(n,j)}(i-(k-1)n+1)I_{k-j-1}^{(n,p,l)}(i)\right)
\]
\[
\fl
=a_{i-(k-1)n}\left(T^{(n,l+k-1)}_{s-1}(i-(k-2)n+1)
-I_{k-1}^{(n,p,l)}(i)    \right.
\]
\[
\left.  -\sum_{j=1}^{k-2}T_{p+(k-j-1)n-2}^{(n,j)}(i-(k-2)n+1)I_{k-j-1}^{(n,p,l)}(i)\right).
\]
For $k=1$ the latter coincides with (\ref{TNh1}). With the help of identities (\ref{t1}), (\ref{t2}) and recurrence relation (\ref{recur}), we can show that the latter is equivalent to the same relation but with $k$ replaced by $k-1$. Therefore, step-by-step, we can show that condition (\ref{t3}) guartantees that (\ref{step}) is valid for any $j\geq 2$. For (\ref{s3}) one can use similar reasonings.

Following remark is in order. It is easy to prove that stationarity condition $S^{(n, l)}_{ln-1}(i+n+1)=S^{(n, l)}_{ln-1}(i)$ mentioned in Introduction is equivalent to condition $S^{(n, l+1)}_{ln-1}(i+1)=S^{(n, l+1)}_{ln-1}(i)$ which is particular case of (\ref{s3}).
This example suggests that Theorem 3 possibly gives all invariant constraints corresponding to submanifolds  ${\cal M}_{n, n+1, 1}\cap{\cal N}_{n,p,l}^d$.

\section{Reductions of Volterra lattice}

\subsection{Invariant constraints for Volterra lattice and its hierarchy}

In this section we apply Theorem 3 in important case of the Volterra lattice
\begin{equation}
a_i^{\prime}=a_i\left(a_{i-1}-a_{i+1}\right).
\label{Vl}
\end{equation}
Evolution equations of Volterra lattice hierarchy look as specialization of (\ref{Bli}), namely
\[
\partial_sa_i=(-1)^sa_i\left(S_{s-1}^s(i-s+2)-S_{s-1}^s(i-s)\right)
\]
with
\[
S^1_0(i)=a_i,\;\;\;
S^2_1(i)=a_iS^1_0(i+1)+a_{i+1}S^1_1(i+1)
\]
\[
=a_ia_{i+1}+a_{i+1}(a_{i+1}+a_{i+2}),
\]
\[
S^3_2(i)=a_iS^2_0(i+1)+a_{i+1}S^2_1(i+1)+a_{i+2}S^2_2(i+1)
\]
\[
=a_ia_{i+1}a_{i+2}+a_{i+1}\left\{a_{i+1}a_{i+2}+a_{i+2}(a_{i+2}+a_{i+3})\right\}
\]
\[
+a_{i+2}\left\{a_{i+1}a_{i+2}+a_{i+2}(a_{i+2}+a_{i+3})+a_{i+3}(a_{i+2}+a_{i+3}+a_{i+4})\right\}
\]
and so on.

Let us restrict ourselves in this section by consideration only reductions of Volterra lattice (\ref{Vl}) generated by conditions of the form
$T^{(1, l+1)}_s(i+1)=T^{(1, l+1)}_s(i)$. Equation (\ref{TNh1}) is specified in this case as \footnote{Here we use simplified notation $T^{(1, l)}_s\equiv T^l_s$.}
\begin{equation}
a_{i+s+l+1}=a_i\frac{T^l_{s-1}(i+2)}{T^l_{s-1}(i+1)}.
\label{disc}
\end{equation}
It should be noted that when $l=0$, the latter is nothing but periodicity condition $a_{i+s+1}=a_i$. For some value $i=i_0$, we denote
$y_1=a_i,\ldots, y_{s+l+1}=a_{i+s+l}$ --- initial data for the discrete equation (\ref{disc}). Then $T_{s-1}^l(i)=T^l_{s-1}(y_1,\ldots, y_{s+l-1})$ is homogeneous polynomial of $l$-th power. In what follows we need in

{\bf Proposition 4.} {\it The function $T_s^l=T^l_s(y_1,\ldots, y_{s+l})$ is invariant with respect to reversing transformation $y_k\mapsto y_{s+l-k+1}$, i.e.,
\begin{equation}
T^l_s(y_{s+l},\ldots, y_1)=T^l_s(y_1,\ldots, y_{s+l}).
\label{property}
\end{equation}
}

{\bf Proof.} By induction with respect to parameter $l$. For $l=1$ the relation (\ref{property}) is evident. Suppose that (\ref{property}) is proved for some value of $l$. Then to prove it for $l+1$ we make of use the identity
\begin{eqnarray}
T_s^{l+1}(y_1,\ldots, y_{s+l+1})&=&\sum_{j=1}^{s-l+1}y_jT_{s-j}^l(y_{j+2},\ldots, y_{s+l+1}) \nonumber \\
                                &=&\sum_{j=1}^{s-l+1}y_{s+l-j+2}T_{s-j}^l(y_1,\ldots, y_{s+l-j}) \nonumber
\end{eqnarray}
which stems from (\ref{tnl1}) and (\ref{tnl}).

Constraining Volterra lattice (\ref{Vl}) by (\ref{disc}) leads to the system of ordinary differential equations
\[
y_1^{\prime}=y_1\left(y_{s+l+1}\frac{T^l_{s-1}(y_1,\ldots , y_{s+l-1})}{T^l_{s-1}(y_2,\ldots , y_{s+l})}-y_2\right),
\]
\begin{equation}
y_k^{\prime}=y_k(y_{k-1}-y_{k+1}),\;\;\; k=2,\ldots, s+l,
\label{ode1}
\end{equation}
\[
y_{s+l+1}^{\prime}=y_{s+l+1}\left(y_{s+l}-y_1\frac{T^l_{s-1}(y_3,\ldots , y_{s+l+1})}{T^l_{s-1}(y_2,\ldots , y_{s+l})}\right).
\]

Compatibility of (\ref{disc}) with (\ref{Vl}) guarantee that the mapping $T : {\bf R}^{s+l+1}\mapsto{\bf R}^{s+l+1}$ given by
\begin{equation}
T(y_k)=y_{k+1},\;\;\;
k=1,\ldots s+l,\;\;\;
T(y_{s+l+1})=y_1\frac{T^l_{s-1}(y_3,\ldots , y_{s+l+1})}{T^l_{s-1}(y_2,\ldots , y_{s+l})}
\label{mapping}
\end{equation}
yields the discrete symmetry.

Observe that the mapping (\ref{mapping}) admits the factorization $T=s_2\circ s_1$, where $s_1$ and $s_2$ are two symmetry transformations acting on variables $\{y_1,\ldots ,y_{s+l+1}, x\}$ as
\[
s_1(y_k)=y_{s+l-k+1},\;\;\;
k=1,\ldots, s+l,
\]
\[
s_1(y_{s+l+1})=y_{s+l+1}\frac{T^l_{s-1}(y_1,\ldots y_{s+l-1})}{T^l_{s-1}(y_2,\ldots y_{s+l})},\;\;\;
s_1(x)=-x
\]
and
\[
s_2(y_k)=y_{s+l-k+2},\;\;\;
k=1,\ldots, s+l+1,\;\;\;
s_2(x)=-x,
\]
respectively. With Proposition 4, one can easily check that $s_1^2=s_2^2=1$. This is evident, of course, for reversing transformation $s_2$. This symmetry is elementary consequence of reversing symmetry of Volterra lattice given by transformation $i\mapsto -i$ and $x\mapsto -x$ and supplemented by appropriate shift $i\mapsto i+\delta$. Having in mind this symmetry one immediately obtains $s_1=s_2\circ T$. It is nontrivial fact only that $s_1^2=1$. The question to be posed is: whether the group of discrete symmetry birational transformations generated by $s_1$ and $s_2$ covers all birational symmetry transformations for the system (\ref{ode1}) or not?

Let us spend some lines to give remarks. It should be noted papers (for example, \cite{adler}, \cite{adler1}, \cite{sklyanin}) where the authors develop a general concept of boundary conditions compatible with higher flows for some integrable lattices. In particular, Adler and Habibullin in \cite{adler} showed that Bogoyavlensky-Volterra finite-dimensional systems associated with a simple Lie algebras \cite{bogoyavlensky} can be derived as a result of imposing special boundary conditions for the Volterra lattice. Our class of constraints yields finite-dimensional systems (\ref{ode1}), including periodic Volterra lattices, which we believe are integrable in Liouville sense.

\subsection{Lax matrices and spectral curves}

Making of use the relation (\ref{ikz}) and Proposition 3, we derive that in terms of KP wave functions, the restriction of discrete KP hierarchy on ${\cal M}_{1,2,1}\cap{\cal N}_{1,l+s+1,l}^1$ is defined by pair of linear equations
\begin{equation}
\fl
z\psi_{i+1}-a_i\psi_{i-1}=z\psi_i\;\;\;\mbox{and}\;\;\;
z^l\psi_{i+l}+\sum_{j=1}^lz^{l-j}T_s^j(i+l-j+1)=w\psi_{i+l+s+1}.
\label{pair}
\end{equation}
Here $w=z^l-\sum_{j\geq 1}I_jz^{-j}$, where $I_j$'s are values of integrals $I_k^{(1,l+s+1,l)}(i)$. Observe that it plays the role of Floquet multiplier. It is evident that the pair (\ref{pair}) is equivalent to equation $L\varphi=0$ on vector-function $\varphi=(\varphi_1,\ldots, \varphi_{l+s+1})$ with some Lax matrix $L$. Here we denote $\varphi_1=\psi_i,\ldots, \varphi_{l+s+1}=\psi_{i+l+s}$. One defines the spectral curve by condition $\det L=0$. It is worth to differ two cases: 1) $s=2g-l$ with $g\geq l$ and 2) $s=2g-l-1$ with $g\geq l+1$. Calculation shows that in the first case the spectral curve is given by algebraic equation
\[
H_0w^2+\left(z^{2g+1}+\sum_{j=1}^gH_jz^{2g+l-j+1}\right)w-z^{2g+l+1}-\sum_{j=1}^lH_jz^{2g+l-j+1}=0
\]
while in the second case it looks like
\[
H_0w^2-\left(z^{2g}+\sum_{j=1}^gH_jz^{2g+l-j}\right)w+z^{2g+l}+\sum_{j=1}^lH_jz^{2g+l-j}=0.
\]
Rational functions $H_j=H_j(y_1,\ldots, y_{s+l+1})$, by construction, are the first integrals of the system (\ref{ode1}).

\section{Conclusion}

In this paper, by using geometric approach, we have shown a broad class of constraints compatible with dynamics defined by INB lattice. All these reductions are defined by some conditions which can be represented as $N$-th order discrete equation
\begin{equation}
a_{i+N}=R(a_i,\ldots, a_{i+N-1})
\label{de}
\end{equation}
with rational right-hand side $R$. Initial data for integration INB lattice constrained by (\ref{de}) is given by vector $(y_1^0,\ldots , y_N^0)\in{\bf R}^N$ which on the one hand gives initial data for discrete equation (\ref{de}) but on the other hand yields initial data for attached autonomous system of ordinary differential equations like (\ref{ode1}), i.e., $y_k^0=y_k(x_0)$. In this connection, the first problem to be addressed is the integration of the systems attached to some constraints in such a way as to present in appropriate form corresponding solutions of INB equation. The second problem which we leave for further investigation is to approve these results on other integrable lattices mentioned in a body of paper.

\section*{Acknowledgments}

This work has been supported by Russian Foundation for Basic Research  grant No. 09-01-00192-a.

\end{document}